\documentclass[
 reprint,
 amsmath,amssymb,
 aps,
 prl,
]{revtex4-1}

\usepackage{graphicx}
\usepackage{dcolumn}
\usepackage{bm}
\usepackage{color}
\usepackage{ulem}
\usepackage{float}
\restylefloat{table}
\begin{document}

\title{What Determines the Fermi Wave Vector of Composite Fermions? }
\date{\today}

\author{D.\ Kamburov}
\author{Yang \ Liu}
\author{M. A. \ Mueed}
\author{M.\ Shayegan}
\author{L. N.\ Pfeiffer}
\author{K. W.\ West}
\author{K. W.\ Baldwin}
\affiliation{Department of Electrical Engineering, Princeton University, Princeton, New Jersey 08544, USA}

\begin{abstract}

Composite fermions (CFs), exotic particles formed by pairing an even number of flux quanta to each electron, provide a fascinating description of phenomena exhibited by interacting two-dimensional electrons at high magnetic fields.  At and near Landau level filling $\nu=1/2$, CFs occupy a Fermi sea and exhibit commensurability effects when subjected to a periodic potential modulation. We observe a pronounced asymmetry in the magnetic field positions of the commensurability resistance minima of CFs with respect to the field at $\nu=1/2$. This unexpected asymmetry is quantitatively consistent with the CFs' Fermi wave vector being determined by the \textit{minority} carriers in the lowest Landau level. Our data indicate a breaking of the particle-hole symmetry for CFs near $\nu=1/2$.
\end{abstract}


\maketitle

Interacting two-dimensional (2D) carriers at low temperatures and in high perpendicular magnetic fields exhibit a remarkable spectrum of many-body states, a hallmark of which is the fractional quantum Hall state (FQHS) \cite{Tsui.1982}. The FQHS is elegantly described through the concept of composite fermions (CFs), weakly interacting quasi-particles formed by pairing an even number of flux quanta with each carrier \cite{Jain.2007,Jain.PRL.1989,Halperin.PRB.1993}. Composite fermions can account for essentially all the FQHSs observed around Landau level filling factor $\nu=1/2$ by treating them as integer QHSs of CFs in a reduced magnetic field given by $B^*=B-B_{1/2}$, where $B_{1/2}$ is the field at $\nu=1/2$ \cite{Jain.2007,Jain.PRL.1989,Halperin.PRB.1993}. Moreover, the CF picture provides two equivalent approaches to understand the FQHSs observed in the range $\nu > 1/2$ and maps them to those seen for $\nu < 1/2$ \cite{Jain.2007}. One approach considers the negative $B^*$ for $B < B_{1/2}$ as a field antiparallel to the external field and maps, e.g., the FQHS at $\nu=2/3$ to the one at 2/5; this predicts FQHSs at effective $B^*$ which are symmetric about $B_{1/2}$. An alternative approach postulates that, instead of having a negative $B^*$, CFs are formed by holes. This approach maps an electron filling factor $\nu$ to a hole filling factor $1-\nu$, e.g., $\nu=3/5$ to 2/5, implying a symmetry between FQHSs in filling factor about $\nu=1/2$. 

Another fundamental property of CFs at and very near $\nu=1/2$ is that they occupy a Fermi sea and therefore possess a Fermi contour \cite{Halperin.PRB.1993,Jain.2007,Willett.PRL.1993,Kang.PRL.1993,Goldman.PRL.1994,Smet.PRL.1996,Smet.1998,Willett.1999,Zwerschke.PRL.1999,Smet.PRL.1999,hCFCOs.Kamburov.2012,Kamburov.PRB.2014}. The CF Fermi contour has been probed in a number of geometrical resonance experiments where the commensurability (or resonance) of the quasi-classical CF cyclotron orbit with the period of a potential modulation in the 2D plane is detected and is used to determine the CF Fermi wave vector  \cite{Kang.PRL.1993,Smet.1998,Willett.1999,Smet.PRL.1999,Zwerschke.PRL.1999,hCFCOs.Kamburov.2012,Kamburov.PRB.2014}. Here we report magnetoresistance data of unprecedented high quality which reveal an unexpected asymmetry in the positions of the CF commensurability features around $\nu=1/2$, both as a function of field and filling factor. On the $B>B_{1/2}$ side, the $B^*$ position of the resonance we observe is consistent with the Fermi wave vector expected for CFs if their density is assumed to be equal to the electron density. For $B<B_{1/2}$, however, the resonance is closer to $B_{1/2}$, implying a smaller Fermi wave vector for the CFs. We can quantitatively account for the asymmetry if we assume that the CF density is equal to the density of the \textit{minority} carriers in the spin-resolved, lowest Landau level (LLL), namely, if the CF density is taken to be the density of electrons when $B>B_{1/2}$ ($\nu<1/2$) but of holes when $B<B_{1/2}$ ($\nu>1/2$). Our results strongly suggest that CFs are formed by pairing up of flux quanta with the \textit{minority} carriers in the LLL. The asymmetry further indicates a breaking of the particle-hole symmetry for CFs near $\nu=1/2$, as it raises the question: Why do electrons, and not holes, determine the CF Fermi wave vector for $B>B_{1/2}$, and vice versa for $B<B_{1/2}$?

\begin{figure*}[t!]
\includegraphics[trim=0 0.0cm 0cm 0cm, clip=true, width=0.90\textwidth]{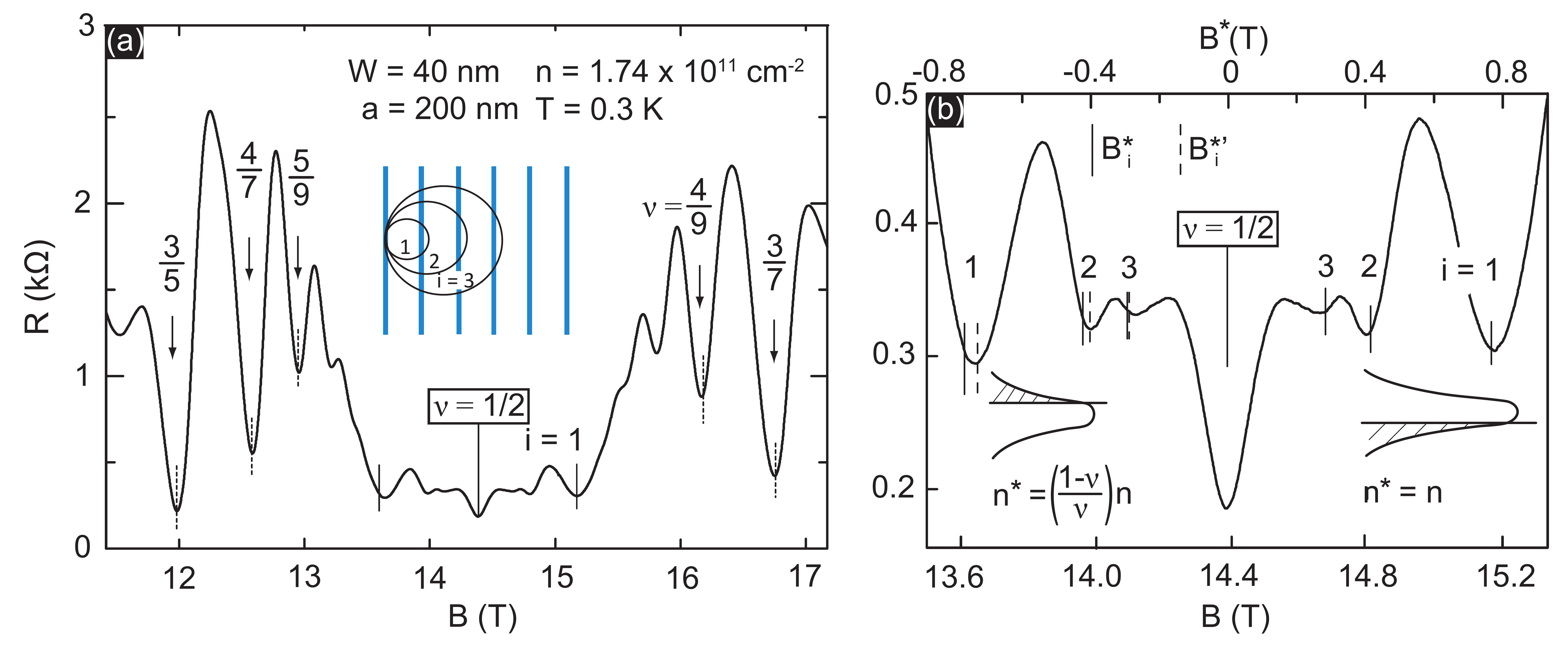}
\caption{\label{fig:Fig1} (color online) (a) Magnetoresistance trace for a 2DES with density $n=1.74 \times 10^{11}$ cm$^{-2}$ and subjected to a periodic potential modulation, exhibiting strong CF commensurability oscillations near $\nu=1/2$. The inset schematically shows the commensurability condition of the quasi-classical CF cyclotron orbits, marked as $i=1$, 2, and 3, with a periodic potential modulation. Dotted vertical lines mark the expected positions of the FQHSs, based on the 2D electron density. (b) The CF commensurability oscillations are shown in greater detail. Vertical solid lines mark the expected positions of the resistance minima when the CF density ($n^*$) is assumed to be equal to the electron density; these positions are symmetric about $\nu=1/2$. If  $n^*$ equals the \textit{minority} density, then the expected positions for the $B<B_{1/2}$ side are those shown with dashed vertical lines. The schematic insets indicate the basis of the CF minority density model which assumes that CFs are formed by the minority carriers in the LLL (hatched parts of the broadened level). }
\end{figure*}

We studied 2D electron and hole systems (2DESs and 2DHSs) confined to symmetric GaAs quantum wells (QWs) grown via molecular beam epitaxy on (001) GaAs substrates. The QW widths are $W=30$ to 50 nm for electrons and 17.5 nm for holes. The carriers are located 131-190 nm under the surface and are flanked on each side by 95-150 nm thick, undoped Al$_{0.24}$Ga$_{0.76}$As barrier layers and $\delta$-doped layers. The $\delta$-doping in the electron and hole samples is Si and C, respectively. The 2D densities are $\simeq 1.2-3.0 \times 10^{11}$ cm$^{-2}$, and the mobilities are $\simeq 10^6 - 10^7$ cm$^2$/Vs. Each sample is covered with a periodic grating of negative electron-beam resist which, through the piezoelectric effect in GaAs, induces a periodic density modulation \cite{Skuras.APL.1997,Endo.PRB.2000,hCFCOs.Kamburov.2012,Kamburov.PRB.2014}. The period of the gratings in our samples range from $a=150$ to 400 nm. We performed experiments in $^3$He refrigerators with base temperatures of $T\simeq$ 0.3 K.

The highlights of our findings are outlined in Fig. 1, where we show the magnetoresistance trace of a 2DES confined to a 40-nm-wide GaAs QW and with a surface grating of 200 nm period. The grating results in a periodic local 2DES density modulation which in turn spatially modulates the effective magnetic field $B^*$ experienced by the CFs in the vicinity of $\nu=1/2$. The modulated magnetic field leads to commensurability oscillations, seen in Fig. 1, flanked by shoulders of higher resistivity \cite{Smet.1998,Willett.1999,Zwerschke.PRL.1999,Smet.PRL.1999,hCFCOs.Kamburov.2012,Kamburov.PRB.2014}. The resistance minima appear at the well-established \textit{magnetic} commensurability condition: $2R^*_C/a=i+1/4$, where $i=1,2,3...$, $a$ is the period of the modulation, $R^*_C=\hbar k^*_F/eB^*$ is the CF cyclotron radius, $k^*_F=\sqrt{4 \pi n^*}$ is the CF Fermi wave vector, and $n^*$ is the CF density \cite{Smet.1998,Willett.1999,Zwerschke.PRL.1999,Smet.PRL.1999,hCFCOs.Kamburov.2012,Kamburov.PRB.2014}. If we assume that $n^*$ is equal to the 2D electron density $n$ on both sides of $\nu=1/2$, the above commensurability condition predicts resistance minima at effective fields $B^*_i= \pm [2\hbar \sqrt{4 \pi n^*}]/[ea(i+1/4)]$ that are symmetric around $B^*=0$ (i.e. around $B_{1/2}$). These $B^*_i$, which are marked with \textit{solid} vertical lines labeled $i=1,2,3$ in Fig. 1, agree with the experimental data for $B>B_{1/2}$, especially for $i=1$ where the deepest minimum is seen.  The commensurability minima for $B<B_{1/2}$, however, appear to the right of the expected values, as clearly seen in Fig. 1(b). 

On the other hand, if we assume that, for $B<B_{1/2}$, CFs in the LLL are formed by the \textit{minority} carriers, i.e., \textit{holes}, the commensurability condition predicts the \textit{dashed} vertical lines in Fig. 1(b), which show better agreement with the experimental data. We elaborate on these observations in the remainder of the paper. We emphasize that the field positions of the FQHSs we observe in the same sample are quite consistent with those expected based on the filling factors and the 2D electron density, as seen by the vertical dotted lines in Fig. 1(a). This is true for the FQHSs observed on \textit{both} sides of $\nu=1/2$.
 
Under the assumption that the density of CFs is equal to the density of the \textit{minority} carriers in the LLL (see insets to Fig. 1(b)), the expected $B^*_i$ for CF commensurability for $B>B_{1/2}$ are the same as before because the minority carrier density equals $n$. For $B<B_{1/2}$, however, according to our assumption, the CF density is equal to the density of $holes$ in the LL: $n^*=\frac{1-\nu}{\nu}n$. Using this $n^*$ in the CF commensurability condition leads to a quadratic equation for the expected positions $B^{* \prime}_i$ of the commensurability minima whose relevant solution can be approximated as $B_i^{* \prime} \simeq B_i^*+B_i^{*2}/B_{1/2}$ \cite{exact_solution}. In this expression we are giving $B_i^{* \prime}$ in terms of $B_i^*$ for the case when $n^*=n$. The expression for $B_i^{* \prime}$ implies that for $B<B_{1/2}$, the minima should be seen closer to $B_{1/2}$ by $\simeq B^{*2}_i/B_{1/2}$. The calculated values of $B_i^{* \prime}$ for $i=1, 2, 3$ are shown in Fig. 1(b) with vertical dashed lines, and are in good agreement with the $B<B_{1/2}$ experimental data. 

\begin{figure}[t!]
\includegraphics[trim=0 0.0cm 0cm 0cm, clip=true, width=0.42\textwidth]{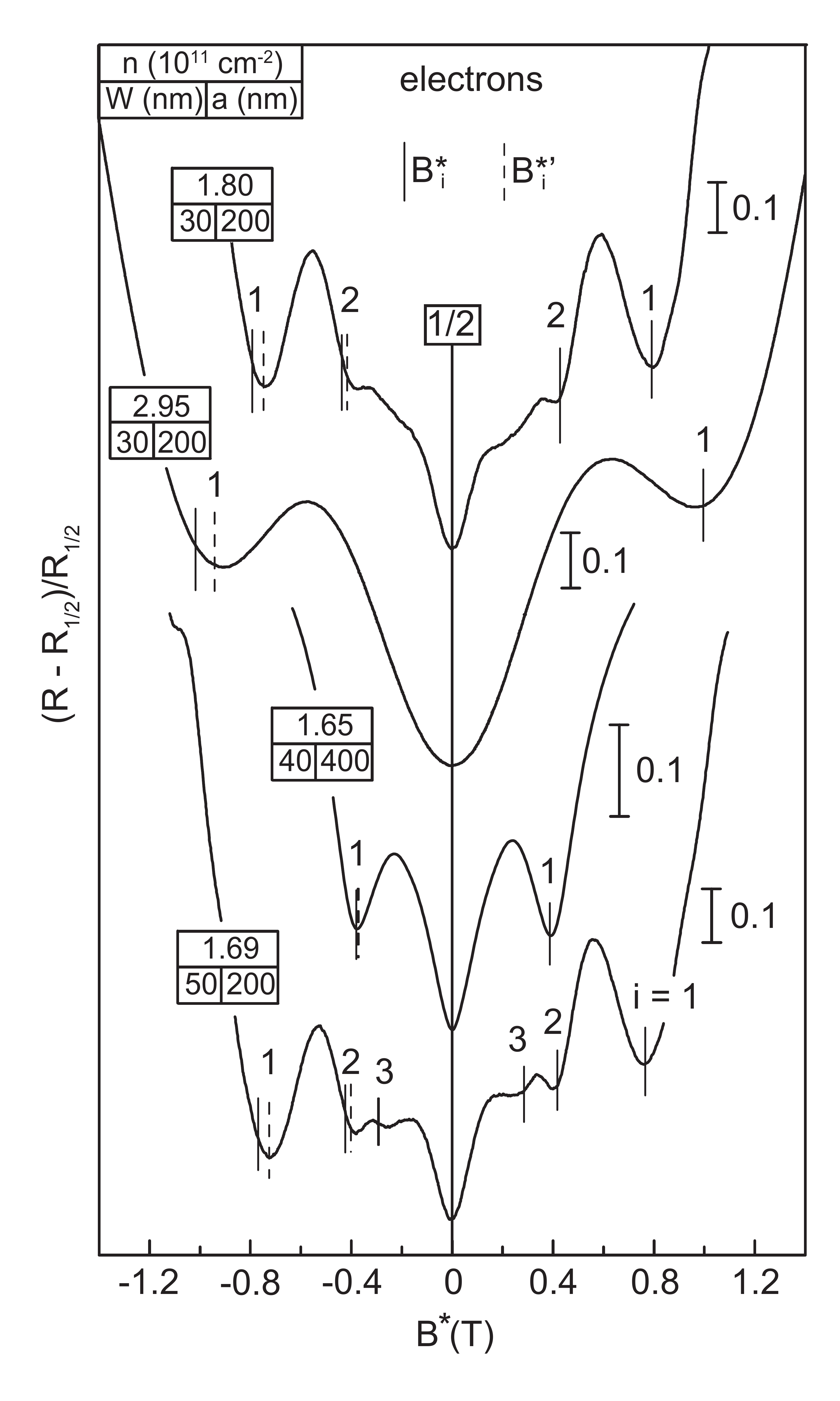}
\caption{\label{fig:Fig1} Magnetoresistance traces near $\nu=1/2$ for four 2D electron samples, normalized to the resistance value $R_{1/2}$ at $B_{1/2}$. Each trace is accompanied with a box whose top row gives the 2D electron density in units of $10^{11}$ cm$^{-2}$, while its bottom row contains the QW width $W$ and the modulation period $a$ in nm. The expected positions of the minima when $n^*=n$ are marked with vertical solid lines which are symmetric about $\nu=1/2$. The dashed lines represent the expected positions from the \textit{minority} CF density model. }
\end{figure}

\begin{figure}[t!]
\includegraphics[trim=0 0.0cm 0cm 0cm, clip=true, width=0.42\textwidth]{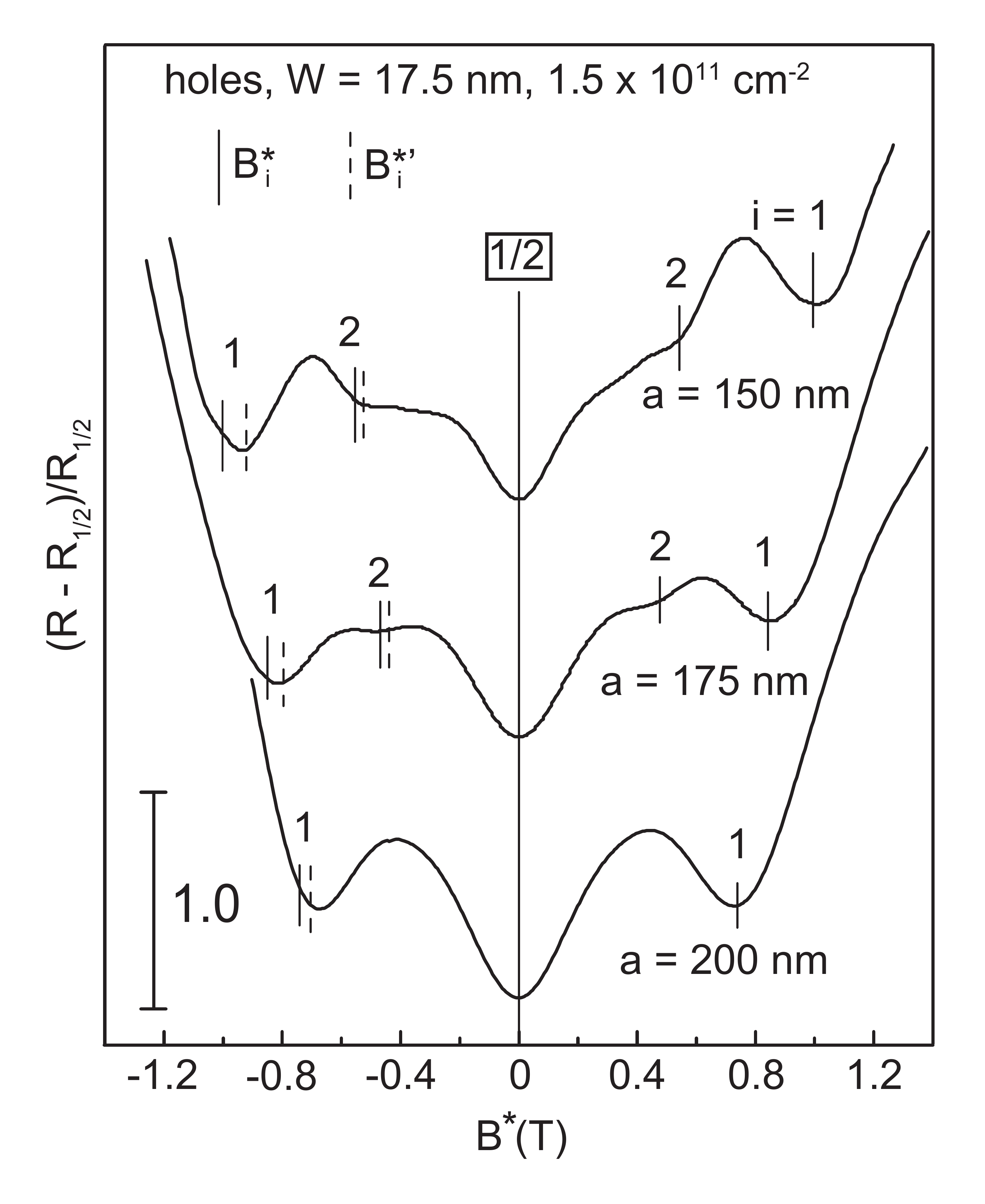}
\caption{\label{fig:Fig1} Magnetoresistance traces, normalized to the resistance value $R_{1/2}$ at $B_{1/2}$ in the vicinity of $\nu=1/2$ for three 2D \textit{hole} samples with $W=17.5$ nm, density $\simeq 1.5 \times 10^{11}$ cm$^{-2}$, and different modulation periods. The expected positions of resistance minima based on $n^*=n$ model are indicated with vertical solid lines which are symmetric about $\nu=1/2$. The dashed lines represent the expected positions from the minority CF density model.}
\end{figure}

\begin{figure*}[t!]
\includegraphics[trim=0 0.0cm 0cm 0cm, clip=true, width=0.9\textwidth]{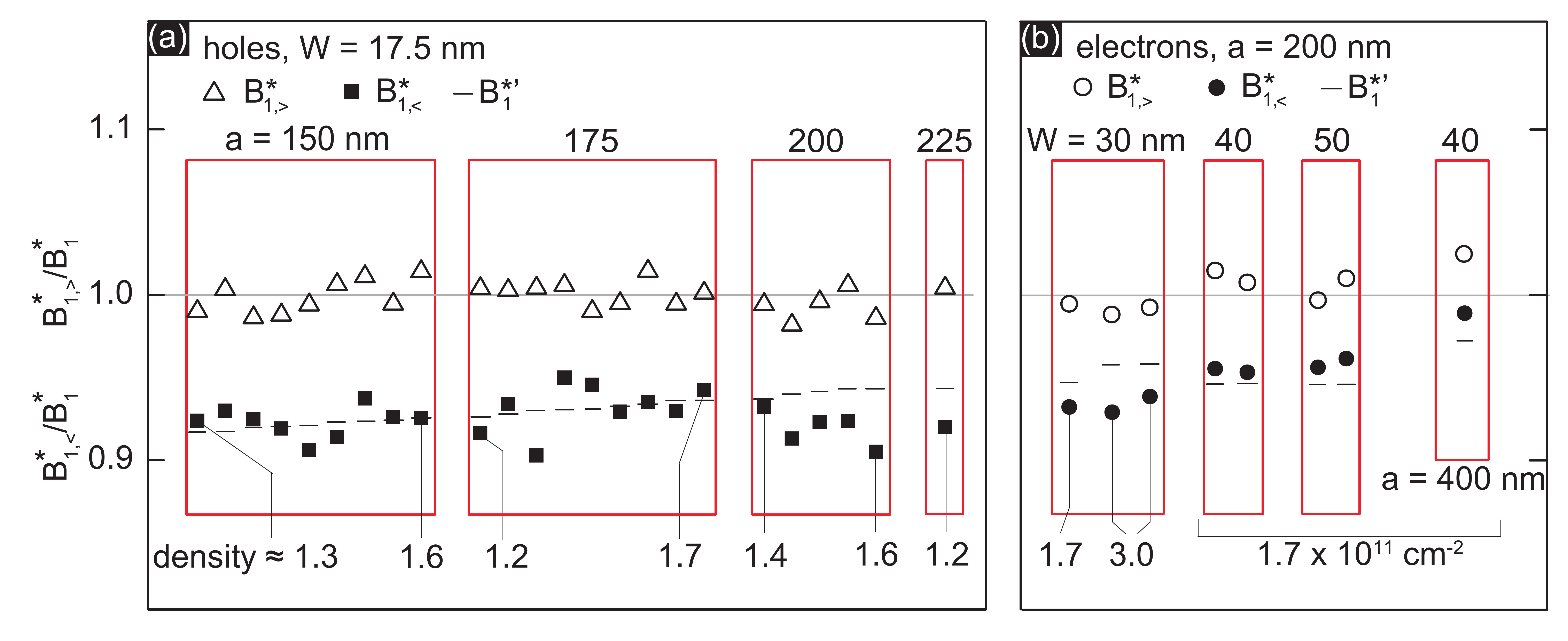}
\caption{\label{fig:Fig1} (color online) Summary of the positions of the observed CF commensurability minima for $i=1$, labeled $B_{1,>}^*$ (open symbols) for $B>B_{1/2}$ and $B_{1,<}^*$ (closed symbols) for $B<B_{1/2}$. Short horizontal lines mark the expected positions $B^{* \prime}_1$ of the minima for $B<B_{1/2}$ based on the minority CF density model. All these fields ($B^*_{1,>}$, $B^*_{1,<}$, and $B_1^{*\prime}$) are normalized to the expected values $B^*_1$ based on the commensurability condition with $n^*=n$. Data for 2DHSs are shown in (a) and are grouped based on the modulation period $a$. The 2D hole density in each group increases from left to right. Data for 2DESs are shown in (b); for all 2DES samples $a=200$ nm except for the last $W=40$ nm QW, where $a=400$ nm. }
\end{figure*}

Having established a possible explanation for the asymmetry in the positions of the CF commensurability resistance minima, we now consider data from samples with different parameters. In Fig. 2, we show data from four 2DES samples. Their density, QW width, and modulation period are given next to each trace, and the expected positions of the CF commensurability minima are indicated with vertical solid and dashed lines. The observed positions of the minima in all traces show an asymmetry, which is well captured by the minority density model. In the $a=400$ nm trace, the minima appear very close to $\nu=1/2$ and are nearly symmetric, as are the predicted positions from both models.

The asymmetry of the CF commensurability minima is not unique to 2DESs. It persists in 2DHSs whose data are shown in Fig. 3. In all samples, the features are asymmetric with respect to $\nu=1/2$, and the asymmetry is captured well by the minority CF density model. Note that for 2DHSs this model implies that CFs are formed by holes for $\nu<1/2$ and by electrons for $\nu>1/2$.

In order to quantify the asymmetry and assess how well it is explained by the minority CF density model, in Fig. 4 we summarize the $i=1$ positions of the observed minima, normalized to the expected value ($B^*_1$) if $n^*=n$ is assumed, for several 2DHSs (a) and 2DESs (b). In all cases the normalized positions of the minima for $B>B_{1/2}$, denoted by $B^*_{1,>}$ (open symbols), are very close to unity, confirming that the $n^*=n$ assumption is consistent with the experimental data. The average value of $B^*_{1,>}/B^*_1$ is 0.998 and the standard deviation is 0.011. On the other hand, the normalized $B<B_{1/2}$ data, denoted by $B^*_{1,<}$ (closed symbols), are smaller than unity for \textit{all} the samples, indicating that an asymmetry exists in all the experimental traces. Figure 4 plots also include the \textit{expected} positions of the minima based on the minority density model ($B^{* \prime}_1$), also normalized to the values if $n^* = n$ were assumed; these $B^{* \prime} _1/B^*_1$ values are shown by short horizontal lines. The experimental data for $B<B_{1/2}$ agree fairly well with the horizontal lines, with a standard deviation of 0.017 \cite{error}. We conclude from Fig. 4 plots that the asymmetry is ubiquitous and its magnitude can be understood reasonably well based on the minority CF density model.

What other factors could cause the asymmetry? One might argue that the asymmetry comes about because the CFs are fully spin polarized when $B>B_{1/2}$ and therefore their Fermi wave vector is consistent with $n^*=n$, but they are only partially polarized when $B<B_{1/2}$ because of the lower $B$. However, our data rule out this possibility. For example, the $B<B_{1/2}$ CF commensurability minimum in the high-density $W=30$ nm 2D electron QW with $n\simeq 3.0 \times 10^{11}$ cm$^{-2}$ is observed at $B\simeq 23.6$ T; this is much higher than $B \simeq 15$ T where the lower density samples show their $B>B_{1/2}$ minimum. Another possibility is that the CFs experience not a pure magnetic potential modulation but rather a mixed magnetic and electrostatic modulation. Such a possibility has been invoked \cite{Zwerschke.PRL.1999,Smet.PRL.1999} to explain an asymmetry observed in the steep resistance rises as one moves farther from $B_{1/2}$, past the commensurability minima. This does not explain the asymmetry we observe in the \textit{positions} of the commensurability minima \cite{electrostatic}. The theoretical calculations, which assume a fixed CF density equal to the electron density, indeed predict minima which are \textit{symmetric} about $B_{1/2}$ \cite{Zwerschke.PRL.1999}. 

To summarize, our data reveal: (1) a persistent asymmetry in the field positions of the commensurability resistance minima of $\nu=1/2$ CFs across many 2D electron and hole samples, (2) the observed minimum for $B>B_{1/2}$ agrees with the position expected for CFs with density equal to the 2D carrier density, but the minimum for $B<B_{1/2}$ is closer to $\nu=1/2$ and its position is consistent with the CF density being equal to the density of \textit{minority} carriers in the LLL. We emphasize that the asymmetry we observe in field positions of the commensurability minima implies an asymmetry, with respect to $\nu=1/2$, of a very similar magnitude in filling factors at which we see these minima. For example, for the data of Fig. 1, we observe commensurability minima at $\nu=0.474$ and 0.527; relative to $\nu=0.500$, this translates to an asymmetry of about 5\% which is similar to the asymmetry in $B^*$ for the same trace (see Fig. 4(b)). Our data thus indicate that the CF commensurability minima are \textit{not} observed at $\nu$ and $1-\nu$, as expected from a simple particle-hole symmetry principle, pointing to a subtle breaking of this symmetry. It is possible that the breaking is caused by the non-idealities, such as disorder, LL mixing and finite layer thickness of real, experimental samples. The combination of LL mixing and three-body interaction, e.g., has been suggested to lead to a lifting of the degeneracy of the Pfaffian and anti-Pfaffian states near $\nu=5/2$ \cite{Levin.PRL.2007,Lee.PRL.2007,Peterson.PRB.2013}. In our case, the LL mixing and the three-body interaction could render a lower ground-state energy for electron-flux CFs compared to the hole-flux CFs for $B > B_{1/2}$, while for $B<B_{1/2}$ the ground-state energies would be reversed. However, we emphasize that, even in \textit{ideal} systems, what determines the Fermi wave vector of CFs is a non-trivial question when one considers that, away from $\nu=1/2$, the electrons and holes in the LLL have unequal densities.

\begin{acknowledgments}
We acknowledge support through the DOE BES (DE-FG02-00-ER45841) for measurements, and the Gordon and Betty Moore Foundation (Grant GBMF2719), Keck Foundation, NSF (ECCS-1001719, DMR-1305691, and MRSEC DMR-0819860) for sample fabrication and characterization. A portion of this work was performed at the National High Magnetic Field Laboratory which is supported by National Science Foundation Cooperative Agreement No. DMR-1157490, the State of Florida and the US Department of Energy. We thank S. Hannahs, T. Murphy, and A. Suslov at NHMFL for valuable technical support during the measurements, and J.K. Jain, M. P. A. Fisher, and B. I. Halperin for illuminating discussions. We also express gratitude to Tokoyama Corporation for supplying the negative e-beam resist \mbox{TEBN-1} used to make the samples.
\end{acknowledgments}


\begin{thebibliography}{99}
\bibitem{exact_solution} The exact solution of the quadratic equation when $n^*=\frac{1-\nu}{\nu}n$ is $B^{* \prime}_i=(B^{*2}_i/B_{1/2}) \Big (1+\sqrt{1+B^2_{1/2}/B^{*2}_i} \Big )$, where $B^*_i$ is the solution when $n^*=n$.
\bibitem{error} In many cases in Fig. 4, when the open symbol falls above (or below) unity, the corresponding closed symbol also falls above (or below) the dashed lines. These pairwise shifts could be explained by possible, small errors in our determination of the field position of $\nu=1/2$.
\end{thebibliography}
\end{document}